\documentclass[aps,prc,twocolumn,showpacs]{revtex4-1}
\usepackage{graphicx}
\usepackage{tikz}
\usepackage{subfigure}

\begin{document}
\title{Measures of azimuthal correlations in relativistic heavy ion collisions}
\author{\small G.L. Li, C.B. Yang and D.M. Zhou}
\affiliation{Institute of Particle Physics \&  Key Laboratory of Quark and Lepton Physics (MOE),
Central China Normal University, Wuhan 430079, People's Republic of China
\small E-mail: cbyang@mail.ccnu.edu.cn
}
\date{\today}

\begin{abstract}
Based on the initial state geometrical symmetry for collisions between two identical heavy ions at high energy, the general form for the one- and two-particle azimuthal
distributions is deduced. Relation between these distributions and the usual flow parameters is
discussed. New measures for the azimuthal correlations are suggested. Some numerical results on
the values of the measures are shown from
an event generator for Au+Au collisions with different colliding centralities  at 200 GeV.

PACS number(s): 25.75.Ag
\end{abstract}
\maketitle

\section{Introduction}

Particle correlation is always one of hot topics of high energy physics for studying
the interactions among the colliding particles. Among various correlations, azimuthal
correlation is crucial in ultra-relativistic $pp$, $pA$ and $AA$ collisions for determining
the event shape and extracting collective information about the produced medium, such as the equation of
state \cite{imp}. In a collision between two identical nuclei with a finite impact
 parameter, the overlap region is approximately an oblong shape in the
  transverse plane. In an intuitive picture, if the interacting
system reaches an approximate local equilibrium and expands according to (viscous) hydrodynamics, the
 geometrical elliptic asymmetry in the initial state will be transformed during the collective expansion
 into an asymmetry in the final state momentum distribution of the detected particles.
 The efficiency of this transformation is sensitive to medium collective properties such as viscosity.
Thus the azimuthal distribution of the produced particles can tell us important information
about the dynamics in the collisions. Since the first data were taken at RHIC, one of the most
important experimental observations \cite{rhic1,rhic2}
has been the azimuthal anisotropy of the detected particles. In particular, the large
value of the so-called elliptic flow observable \cite{volo}, which indicates strong collective behavior
of the produced system, has been one of the most important and most frequently studied measurements.
The measurement of the collective flow parameters provided one of the strongest pieces of evidence for
the creation of a strongly-coupled, low-viscous
QGP medium in these collisions. With other measurements, those parameters have the potential to provide tight
 constraints on models on extracting precise quantitative properties of the QGP, as well as to shed
 light on the non-equilibrium QCD dynamics of the initial stage of the collision, which are  poorly understood
 up to now.

Because of fluctuations in the colliding system's evolution and particle production processes and due to
the lack of solid theoretical ground for calculating the correlation functions from first principles,
our understanding of the correlations is quite limited up to now to some model analysis. Related
with the azimuthal correlations, azimuthal flow effect is one of the centers in both experimental
and theoretical studies of particle production
mechanism and interactions of various components in the system.

In this paper, more measures for the collective flow effect are suggested from the two-particle azimuthal
correlations. These measures are proposed based on some initial geometrical symmetry and thus are very
general, independent of the interactions during the evolution of the produced partons and hadrons.
Relation with the usual flow measurement is discussed also.
 Though the discussion focuses mainly on correlations in the mid-rapidity region,
 extension to other rapidity region is straightforward.
The measures discussed here may be used to constrain
theoretical models for relativistic heavy ion collisions.
By using an event generator, some numerical results are obtained for Au+Au collisions at
$\sqrt{s_{NN}}=200$ GeV with different centralities.

This paper is organized as follows. In section II the most general form for the one- and two-particle distributions are deduced
based on the (event averaged) geometrical symmetry. Section III is for numerical results for the coefficients in the distributions
from a Monte Carlo model. Section IV is for conclusions and discussions.

\section{General form of one- and two-particle azimuthal distributions}
\subsection{For the case with all particles in the same kinematic region}
Let us first study particles in some prefixed kinematic region.
Consider an event with $\tilde{N}$ final state particles in that kinematic region, for example with rapidity $|y|<0.5$ and
$p_T<1$GeV$/c$, with azimuthal angles $\varphi_1, \varphi_2,\cdots, \varphi_{\tilde{N}}$ in the reaction plane of the colliding system. The one- and two-particle azimuthal
distributions for the event can be written as
\begin{eqnarray}
\widetilde{\rho}_1(\Phi)&=&\sum_{i=1}^{\tilde{N}}\delta(\Phi-\varphi_i)\ ,\\
\widetilde{\rho}_2(\Phi_1,\Phi_2)&=&\sum_{i\neq j=1}^{\tilde{N}} \delta(\Phi_1-\varphi_i)
\delta(\Phi_2-\varphi_j)\ .
\end{eqnarray}
In the above expressions, a tilde is used to represent distributions fluctuating from event to
event, including variations of multiplicity and azimuthal angles of particles in an event.
These two distributions are normalized to ${\tilde{N}}$ and $\tilde{N}(\tilde{N}-1)$, respectively,
\begin{equation}
\begin{array}{cc}
      & \int_0^{2\pi}d\Phi \widetilde{\rho}_1=\tilde{N}\ , \\
      & \int_0^{2\pi} d\Phi_1d\Phi_2\widetilde{\rho}_2(\Phi_1,\Phi_2)=\tilde{N}(\tilde{N}-1)\ .
\end{array}
\label{norm}
 \end{equation}
 From the definition of $\widetilde{\rho}_2$, one can observe the exchange symmetry
$\widetilde{\rho}_2(\Phi_1,\Phi_2)=\widetilde{\rho}_2(\Phi_2,\Phi_1)$.

One can rewrite the above distributions as sums of infinite set of sine and cosine functions
\begin{widetext}
\begin{eqnarray}
\widetilde{\rho_1}(\Phi)&=&\sum_{k=0}^{\infty}(\widetilde{s}_k \sin(k\Phi)+\widetilde{c}_k\cos(k\Phi))\ ,\label{de1}\\
\widetilde{\rho_2}(\Phi_1,\Phi_2)&=&\sum_{k,l=0}^{\infty}(\widetilde{s}_{k,l}\sin(k\Phi_1)\sin(l\Phi_2)+
\widetilde{d}_{k,l}(\sin(k\Phi_1)\cos(l\Phi_2)+\cos(k\Phi_1)\sin(l\Phi_2))+
\widetilde{c}_{k,l}\cos(k\Phi_1)\cos(l\Phi_2))\ ,\label{de2}
\end{eqnarray}
\end{widetext}
with $\widetilde{s}_{k,l}=\widetilde{s}_{l,k},\ \widetilde{d}_{k,l}=\widetilde{d}_{l,k},\
\widetilde{c}_{k,l}=\widetilde{c}_{l,k}$.
From the normalization conditions in Eq.(\ref{norm}), one gets $\widetilde{c}_0=\tilde{N}/(2\pi),
\widetilde{c}_{0, 0}=\tilde{N}(\tilde{N}-1)/(4\pi^2)$.

Theoretically and experimentally, one can investigate the distributions after averaging over many
events and obtain expressions for the averaged one- and two-particle azimuthal distributions in the same
format as in last two equations, without tilde on all coefficients of Fourier terms
\begin{widetext}
\begin{eqnarray}
{\rho_1}(\Phi)&=&\sum_{k=0}^{\infty}({s_k} \sin(k\Phi)+{c_k}\cos(k\Phi))\ ,\label{de3}\\
{\rho_2}(\Phi_1,\Phi_2)&=&\sum_{k,l=0}^{\infty}({s_{k,l}}\sin(k\Phi_1)\sin(l\Phi_2)+
{d_{k,l}}(\sin(k\Phi_1)\cos(l\Phi_2)+\cos(k\Phi_1)\sin(l\Phi_2))+
{c_{k,l}}\cos(k\Phi_1)\cos(l\Phi_2))\ .\label{de4}
\end{eqnarray}
\end{widetext}
The normalization of these two distributions reads $c_0=\langle \tilde{N}\rangle, c_{0, 0}=\langle
\tilde{N}(\tilde{N}-1)\rangle$, with $\langle\cdots\rangle$ for the average over many events.

\begin{figure}[tbph]
\includegraphics[width=0.43\textwidth]{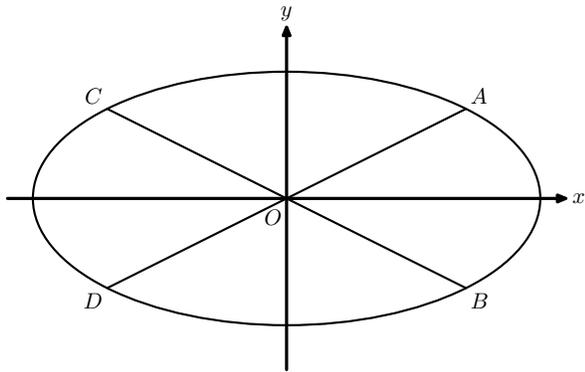}
\caption{A schematic illustration for the geometric symmetry of the colliding system in the  transverse
plane at mid-rapidity or in the whole rapidity region. The $x$ axis is along the direction of the impact parameter.}
\label{fig1}
\end{figure}

For collisions between two identical nuclei, Au+Au for example, the interaction region
is approximately almond in the transverse plane after averaging over many events
when the reaction plane is rotated to the same $xy$ plane, as schematically shown in Fig. \ref{fig1}, therefore the particle
 distribution at central rapidity  or in the whole rapidity region should satisfy the same symmetry.
Due to the initial geometrical symmetry, properties at directions $OA, OB, OC$ and $OD$ must be the same
if $\angle AOX$ is equal to $\angle BOX$. Thus one can demand the above distributions
satisfy the following conditions
\begin{eqnarray}
\rho_1(\Phi)&=&\rho_1(\pi-\Phi)=\rho_1(-\Phi)\, \label{sy1}\\
\rho_2(\Phi_1,\Phi_2)&=&\rho_2(-\Phi_1,-\Phi_2)\nonumber\\
&=&\rho_2(\pi-\Phi_1,\pi-\Phi_2)\ .\label{sy2}
\end{eqnarray}

From these demands, one gets, from Eqs. (\ref{de3}) and (\ref{de4})
\begin{eqnarray*}
&& s_k=0,\ \ c_{2k+1}=0,\ \ d_{k,l}=0,\\
&& c_{2k,2l+1}=c_{2k+1,2l}=0, \ \ s_{2k,2l+1}=s_{2k+1,2l}=0.
\end{eqnarray*}
Therefore, $\rho_1$ and $\rho_2$ can be rewritten as
\begin{eqnarray}
\rho_1(\Phi)&=&c_0+\sum_{k=1}^\infty c_{2k} \cos 2k\Phi\ ,\label{dis1}\\
\rho_2(\Phi_1,\Phi_2)&=& \Delta_{0,0}+{\sum_{k,l}}^\prime
\left[2\Delta^+_{k,l} \cos(k\Phi_1+l\Phi_2)\right.\nonumber\\
&& \left. +2\Delta^-_{k,l}\cos(k\Phi_1-l\Phi_2)\right],\label{dis2}
\end{eqnarray}
where $\sum_{k,l}^\prime$ means that the summation should be performed for all
non-negative integers $k$ and $l$ with $k+l$ an even number larger than zero,
or in other words, non-negative $k$ and $l$ should be both odd or
even in every term and at least one of them is  positive. Eq. (\ref{dis1}) is the usual expression for the inclusive
azimuthal distribution, and the coefficients $c_{2k}$ are related to the
flow measures $v_{2k}$ by
$v_{2k}=c_{2k}/(2c_0)$ for $k=1, 2, \cdots$. Because of the exchange symmetry
$\rho_2(\Phi_1,\Phi_2)=\rho_2(\Phi_2,\Phi_1)$ for pairs of particles in the same kinematic region,
$\Delta^\pm_{k,l}$ are symmetric under exchange of their indices
\begin{equation}
\Delta^\pm_{k,l}=\Delta^\pm_{l,k}\ .
\end{equation}
Those $\Delta^\pm_{k,l}$ contain more information on the azimuthal distributions of the final state particles than the flow parameters.
Thus they are new measures for collective effect as
\begin{eqnarray}
\Delta_{2n,0}/\Delta_{0,0} &=& \langle \cos(2n\Phi_1)\rangle=v_{2n}\ ,\label{de5}\\
\Delta^\pm_{k,l}/\Delta_{0,0} &=&\langle\cos(k\Phi_1\pm l\Phi_2)\rangle\ .\label{de6}
\end{eqnarray}
Here $\langle\cdots\rangle$ means average over the corresponding distributions.

It is interesting to note that $\Delta_{11}^+/\Delta_{00}=\langle\cos(\Phi_1+\Phi_2)\rangle$
has been suggested in \cite{vol} as a quantity to detect the presence of $CP$-odd domains
\cite{cp} in the deconfined  QCD vacuum. Also, the new expressions, Eqs. (10) and (11) are in agreement with the latest
results \cite{brav} on the absence of directed flow in the central rapidity region.

For azimuthal distributions in the forward/backward rapidity region or for collisions between two
non-identical nuclei, $\rho_1(\Phi)$ and $\rho_2(\Phi_1,\Phi_2)$ satisfy the symmetry conditions
\begin{eqnarray}
\rho_1(\Phi)&=& \rho_1(-\Phi)\neq \rho_1(\pi+\Phi),\\
\rho_2(\Phi_1,\Phi_2)&=&\rho_2(-\Phi_1,-\Phi_2)\nonumber\\
&\neq& \rho_2(\pi+\Phi_1,\pi+\Phi_2)\ ,
\end{eqnarray}
then there are odd harmonic terms $\cos(2k+1)\Phi$ in the Fourier expansions of $\rho_1$
and terms with odd $k+l$ in $\rho_2$. This can be concluded from results in \cite{brav}.

Very often, one needs the distribution $P(\Delta\Phi)$ of $\Delta\Phi=\Phi_1-\Phi_2$, which can be obtained
 without determination of the reaction plane. From the expression for $\rho_2$, one readily gets
\begin{eqnarray}
P(\Delta\Phi)&=&\int d\Phi_1d\Phi_2 \rho_2(\Phi_1,\Phi_2)\delta(\Delta\Phi-\Phi_1+\Phi_2)\nonumber\\
  &=& 2\pi\left(\Delta_{0,0}+\sum_{k=1}^\infty 2\Delta^-_{k,k}\cos(k\Delta\Phi)\right)\ .
  \label{dp}
  \end{eqnarray}
  In this expression, $k$ can be odd and even. From this expression, one can see that generally
  $P(\Delta\Phi=0)\neq P(\Delta\Phi=\pi)$, as observed experimentally \cite{back}.
  For high $p_T$ particles such phenomena have been explained
in the framework of jet quenching \cite{quench}. By comparing coefficients in $\rho_2(\Phi_1,\Phi_2)$
 and $P(\Delta\Phi)$, one can see easily that some correlation information in
 $\rho_2(\Phi_1,\Phi_2)$ is washed out in obtaining the $\Delta\Phi$ distribution from
 $\rho_2(\Phi_1,\Phi_2)$. However, the above expression is different from the usual
  parametrization for $P(\Delta\Phi)$ used by experimentalists, where $P(\Delta\Phi)$ was parameterized
  by only terms of $\cos 2k\Delta\Phi$ with $k$ an integer number. In our new expression, $\cos (2k+1)
  \Delta\Phi$ terms are present. Because of the presence of those odd terms in $P(\Delta\Phi)$,
  $P(\Delta\Phi)\neq P(\pi-\Delta\Phi),\ P(\Delta\Phi)\neq P(\pi+\Delta\Phi)$. These two consequences
  can be tested easily in experiments.

  If there were no azimuthal correlations among the produced particles,  the same
  information about flow would be contained in $\rho_2(\Phi_1,\Phi_2)$ and $\rho_1(\Phi_1)\rho_1(\Phi_2)$, then $\rho_2$ would be  factorized.   Such a factorization has been used experimentally to
 determine the elliptic flow coefficient $v_2$ \cite{rhic1}.
Such a factorization may be expected when the soft particles are emitted
from thermalized medium independently.
 If this factorization is true, one can get
  \begin{equation}
  \Delta^\pm_{2k,2l}=c_{2k}c_{2l}\ , \Delta^\pm_{2k+1,2l+1}=0\ .
  \end{equation}
  In particular, for $k=l=0$, the above condition reads $\langle N(N-1)\rangle=\langle N\rangle^2$,
thus the multiplicity fluctuation must be of Poissonian.
Under the above condition, the $\Delta\Phi$ distribution could be obtained then from products of $\rho_1$'s as
\begin{equation}
P(\Delta\Phi)=2\pi\left(c^2_0+\sum_{k=1}^\infty 2c^2_{2k} \cos(2k\Delta\Phi)\right)\ ,
\label{fact}
\end{equation}
with the coefficient ratios $c^2_{2k}/c^2_0=v^2_{2k}$. The above equation has been used
in \cite{rhic1} to measure the flow coefficients for Au+Au collisions at $\sqrt{s_{NN}}=130$ GeV.
However the validity of the above expression has never been proved. In fact, some experimental data have
shown that some  correlation variables are not the same at $\Delta\Phi=0$ and $\pi$, as in the $\Delta\Phi$ dependence of the joint autocorrelations at $\Delta\eta=0$ in \cite{STAR1}, in the charged di-hadron
distribution in the $\Delta\eta-\Delta\Phi$ plane in \cite{STAR2}, and in the correlation structure shown in
Fig. 3 in \cite{STAR3}. From Fig. 3 in \cite{STAR3}, one can see clearly that the correlation structure
is not the same at $\Delta\Phi=0$ and $\pi$. From the same figure, one can see also that even for $p+p$
 collisions  the correlation structure from PYTHIA simulation can not be well described by expressions
 with only $\cos 2k\Delta\Phi$ terms, like Eq. (19).

 By comparing Eq. (\ref{fact}) with Eq. (\ref{dp}), one can see that terms with
  odd $k$ in
 the Fourier expansion of $P(\Delta\Phi)$, Eq. (\ref{dp}), are absent in the factorization scheme.
 Thus the presence of $\cos((2k+1)\Delta\Phi)$ terms in the $\Delta\Phi$ distribution disvalidates the
 factorization of the two-particle azimuthal angle distribution.

\subsection{For the case with particles within different kinematic regions}
Now we turn to the measure of correlations between two sets of particles selected from two different kinematic
 regions. As an example, let one set of particles come from low $p_T\ (\le 1 {\rm GeV}/c)$ region (soft particles), another from high $p_T\ (\ge 2{\rm GeV}/c)$ region (hard particles).
The high $p_T$ particles are frequently called triggers which can be neutral pions, photons,  etc.
 Then one should use two one-particle distributions for the soft and hard particles, respectively.
 The azimuthal distributions are
  \begin{eqnarray}
  \widetilde{\rho}_{1s}(\Phi_1)&=&\sum_i^{\tilde{N}_s} \delta(\Phi_1-\varphi_{1i})\ , \\ \widetilde{\rho}_{1h}(\Phi_2)&=&\sum_i^{\tilde{N}_h}
  \delta(\Phi_2-\varphi_{2i})\ ,\\
  \widetilde{\rho}_2(\Phi_1,\Phi_2)&=& \sum_{i=1}^{\tilde{N}_s}\sum_{j=1}^{\tilde{N}_h}\delta(\Phi_1-\varphi_{1i})\delta(\Phi_2-\varphi_{2j})\ ,
  \end{eqnarray}
where $\tilde{N}_s$ and $\tilde{N}_h$ are multiplicities of the soft and hard particles in an event.

One can perform Fourier decomposition to the distributions and obtain expressions similar to Eqs.
(\ref{de1},\ref{de2}) with coefficients fluctuating from event to event. After averaging over many
events, smooth distributions similar to Eqs. (\ref{de3},\ref{de4}) can also be obtained at mid-rapidity  as
\begin{eqnarray}
\rho_{1s}(\Phi_1)&=&\frac{\langle N_s\rangle}{2\pi}\left(1+\sum_{k=1}^\infty 2v^s_{2k}\cos(2k\Phi_1)\right)\ ,\\
\rho_{1h}(\Phi_2)&=&\frac{\langle N_h\rangle}{2\pi}\left(1+\sum_{k=1}^\infty 2v^h_{2k}\cos(2k\Phi_2)\right)\ ,\\
\rho_2(\Phi_1, \Phi_2)&=&\frac{\langle N_sN_h\rangle}{4\pi^2}+{\sum_{k+l>0}}
[2\Delta^+_{k,l}\cos(k\Phi_1+l\Phi_2)\nonumber\\
&&  +2\Delta^-_{k,l}\cos(k\Phi_1-l\Phi_2)])\ .
\label{dis3}
\end{eqnarray}
In the above equations, $\langle \cdots\rangle$ represents average over many events.

Since these two sets
of particles are from two different kinematic regions, no exchange symmetry can be expected for $\rho_2$.
Because of the geometrical symmetry of the colliding system, the symmetry properties depicted  in
Eqs. (\ref{sy1},\ref{sy2}) are still valid at mid-rapidity or for the whole rapidity region.
Because of the absence
of the exchange symmetry, $\rho_2(\Phi_1,\Phi_2)\neq \rho_2(\Phi_2,\Phi_1)$,
the coefficients in $\rho_2(\Phi_1,\Phi_2)$ are not symmetric under the exchange of their indices,
$\Delta^\pm_{k,l}\neq \Delta^\pm_{l,k}$. Equations similar to Eqs. (\ref{de5}) and (\ref{de6}) can
be written for the soft and hard particles.

Experimentally, the reaction plane in a nucleus-nucleus collision is not known and must be determined
from the produced particles. Of course, the determined $x-$axis has an angle $\Phi_{\rm rec}$ relative
to the true $x-$axis. Then for each final state particle, the azimuthal angle detected can be related to the
true value $\phi_i$ by $\phi_{i, \ \rm exp}=\phi_i-\Phi_{\rm rec}$. From this relation, the averages of $\cos(k\Phi_1\pm\Phi_2)$
from experiments can be written as
\begin{equation}
\langle\cos(k\Phi_1\pm\Phi_2)\rangle_{\rm exp}=\langle\cos(k\Phi_1\pm l\Phi_2-(k\pm l)\Phi_{\rm rec})\rangle\ .
\end{equation}
If one assumes that the experimentally determined reaction plane is distributed symmetrically around the true one and accurate enough (see Ref.{\cite{PHOB}} for the state of the art on determining reaction plane and the flow coefficients), one has
\begin{eqnarray*}
\langle\sin((k\pm l)\Phi_{\rm rec})\rangle=0\ ,
\langle\cos((k\pm l)\Phi_{\rm rec})\rangle>0\ .
\end{eqnarray*}
Then
\begin{equation}
\langle\cos(k\Phi_1\pm\Phi_2)\rangle_{\rm exp}\propto\langle\cos(k\Phi_1\pm l\Phi_2)\rangle\ .
\end{equation}
The two-particle correlation function has the same form as from our theoretical consideration.
Thus one can observe similar two-particle correlation distributions from experimental data. The same conclusion can
be claimed for the single-particle azimuthal distributions Eqs. (10), (23) and (24).

\section{Numerical results from a transport model}

The above discussions are based on the geometrical symmetry properties of the colliding heavy ion systems, thus
should be valid for both Au+Au and Pb+Pb collisions at all colliding energies and centralities.
Before real experimental data is available to test the above conclusions, one can use a Monte Carlo
event generator for producing ``the experimental data" and calculating the coefficients in the relevant expressions.

For the purpose of generating ``experimental data'',
a transport model, AMPT \cite{am1}, is used to generate Au+Au collision events at
$\sqrt{s_{NN}}$=200 GeV with different centralities.
The AMPT model is a multi-phase transport  model \cite{am1}, which is constructed to describe nuclear collisions. It includes initial partonic and final hadronic interactions, and the transition between these two phases. The model consists of four main processes: the initial conditions, partonic interactions, conversion  from partonic matter to the hadronic matter and the hadronic rescattering. There are two kinds of AMPT model ,the default AMPT and the AMPT model with string melting. HIJING model \cite{am2} provides the initial momentum and spatial distribution  of minijet partons and soft string excitations. Parton scattering in the AMPT model is implemented by using the ZPC model \cite{am3}. In the default version of AMPT model, partons are recombined with their parent strings when they stop interacting,and Lund string fragmentation model is used to convert the resulting strings to hadrons. In the string melting version, a quark coalescence model is used to combine partons into hadrons. The dynamics of hadronic matter is modelled by  a relativistic hadronic transport model (ART)\cite{am4} .

The string melting version of AMPT model was used in our study, since it can give a reasonable description of  flow of Au+Au collisions \cite{am5,am6,am7,am8}. With the increase of  impact parameter, the number of produced particles in an event decreases, thus more events need to be generated for the average of quantities to reduce the statistical fluctuation. In our calculation, about four hundred thousand to four million events are generated from the AMPT model for different centralities (from $10\%$ to $90\%$) for the analysis in the following. In the analysis, only charged hadrons within rapidity region $|\eta|<1.0$ are considered. Among those particles, hadrons with $p_T<1 {\rm GeV}/c$ are referred as soft particles, and those with $p_T>2{\rm GeV}/c$ as hard particles. From the azimuthal angles of soft and hard particles one can calculate, for pairs of soft-soft, hard-hard and soft-hard in every event, contributions to the $\Delta^\pm_{k, l}$ for any chosen non-negative integer $k, l$, and
then obtain the coefficients discussed in the last section.

\begin{figure}[tbph]
\includegraphics[width=0.5\textwidth]{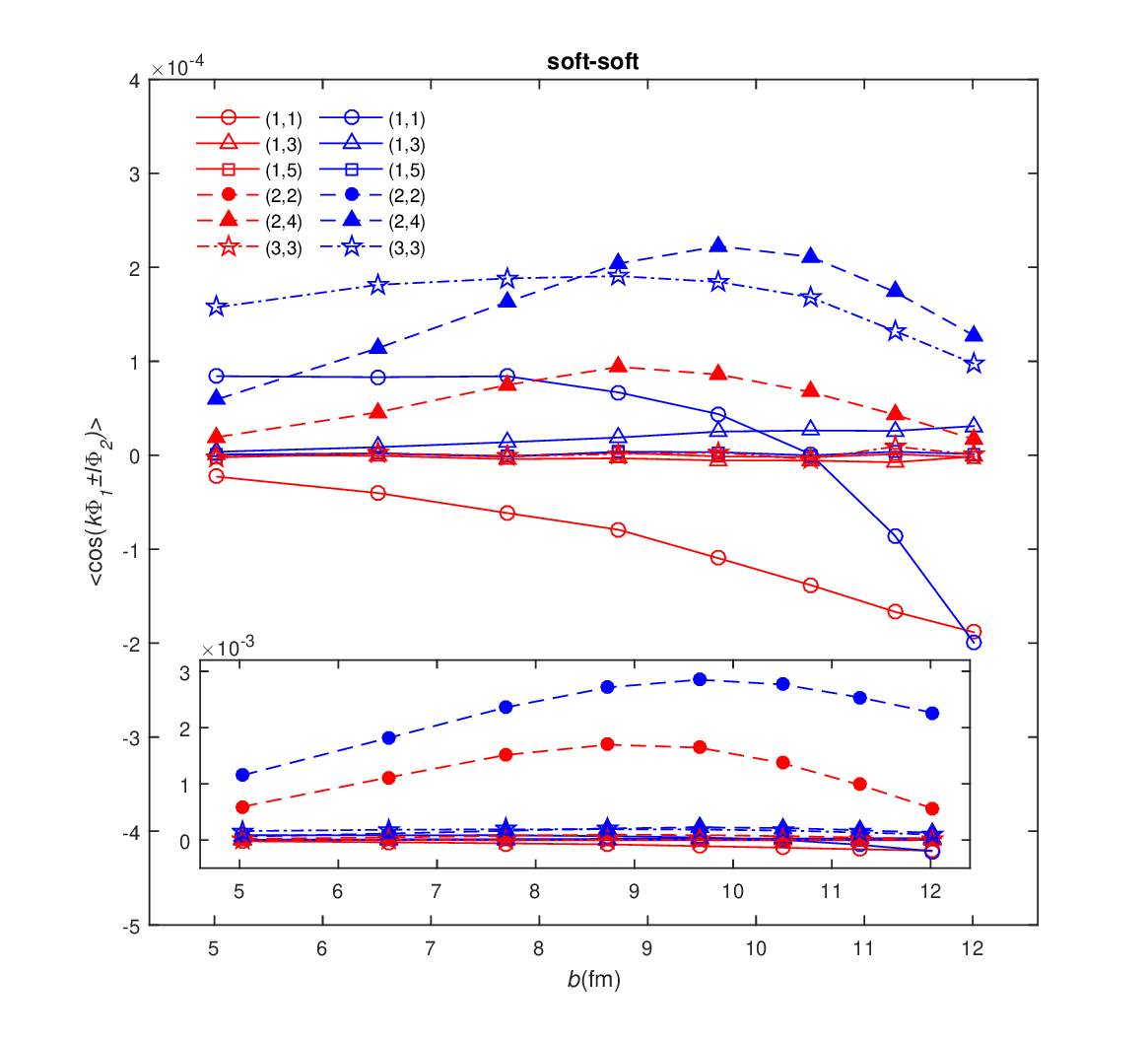}
\caption{(Color online) Centrality dependence of $\langle\cos(k\Phi_1\pm l\Phi_2)\rangle$ for different combinations of $(k,l)$ in Eq. (\ref{dis2}) for all final state soft hadrons with $p_T$ smaller than 1 GeV/$c$ for Au+Au collisions at $\sqrt{s_{NN}}.$
=200 GeV.  Red lines and  markers are for  $\langle\cos(k\Phi_1+ l\Phi_2)\rangle$, blue lines and markers for $\langle\cos(k\Phi_1- l\Phi_2)\rangle$. }
\label{fig2}
\end{figure}

\begin{figure}[tbph]
\includegraphics[width=0.5\textwidth]{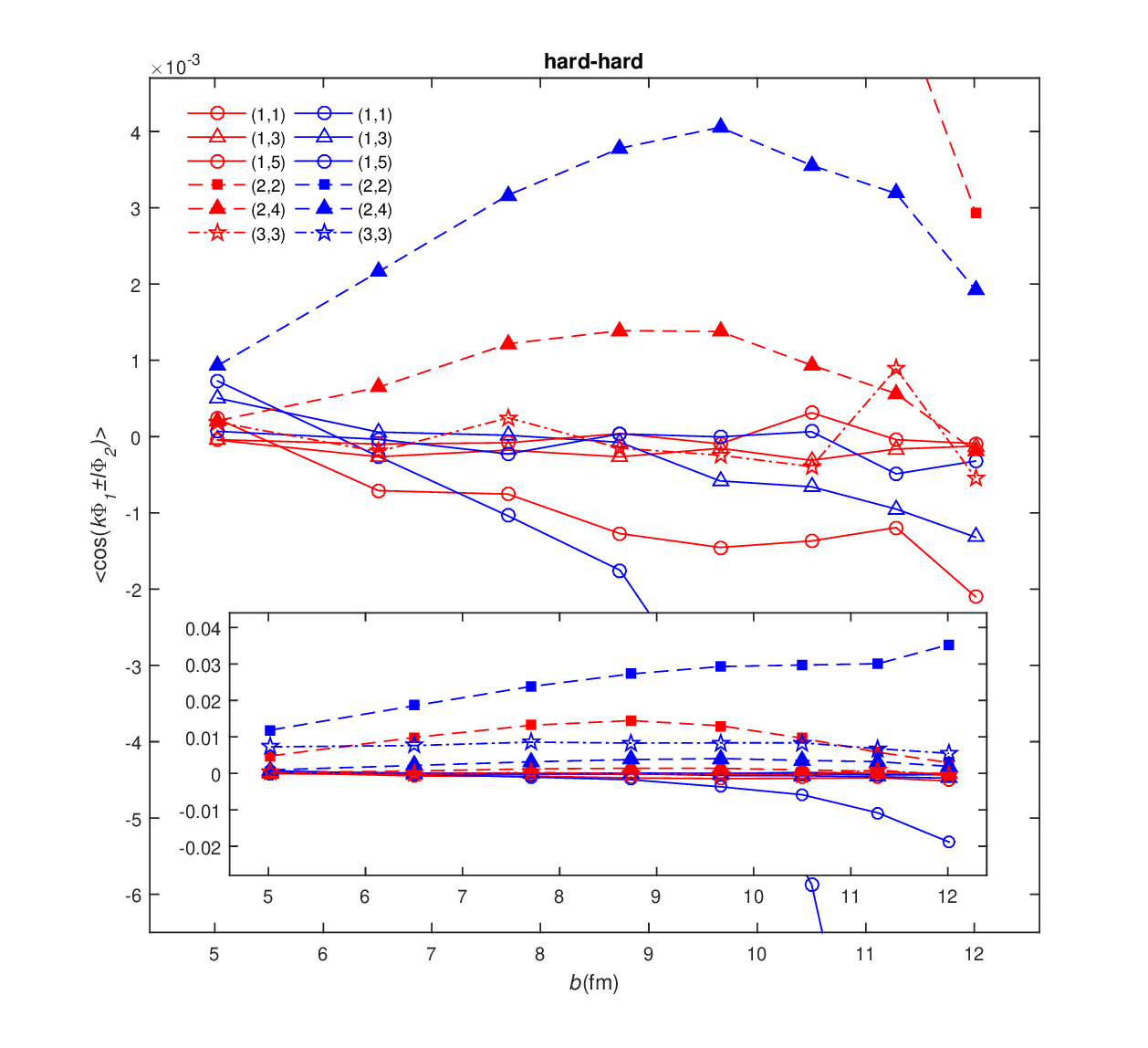}
\caption{(Color online) Centrality dependence of $\langle\cos(k\Phi_1\pm l\Phi_2)\rangle$ for different combinations of $(k,l)$ in Eq. (\ref{dis2}) for all final state hard hadrons with $p_T$ larger than 2 GeV/$c$ for Au+Au collisions at $\sqrt{s_{NN}}.$
=200 GeV.  Red lines and  markers are for $\langle\cos(k\Phi_1+ l\Phi_2)\rangle$, blue lines and  markers for $\langle\cos(k\Phi_1- l\Phi_2)\rangle$. }
\label{fig3}
\end{figure}

\begin{figure}
\centering
\subfigure{ 
\includegraphics[width=0.5\textwidth]{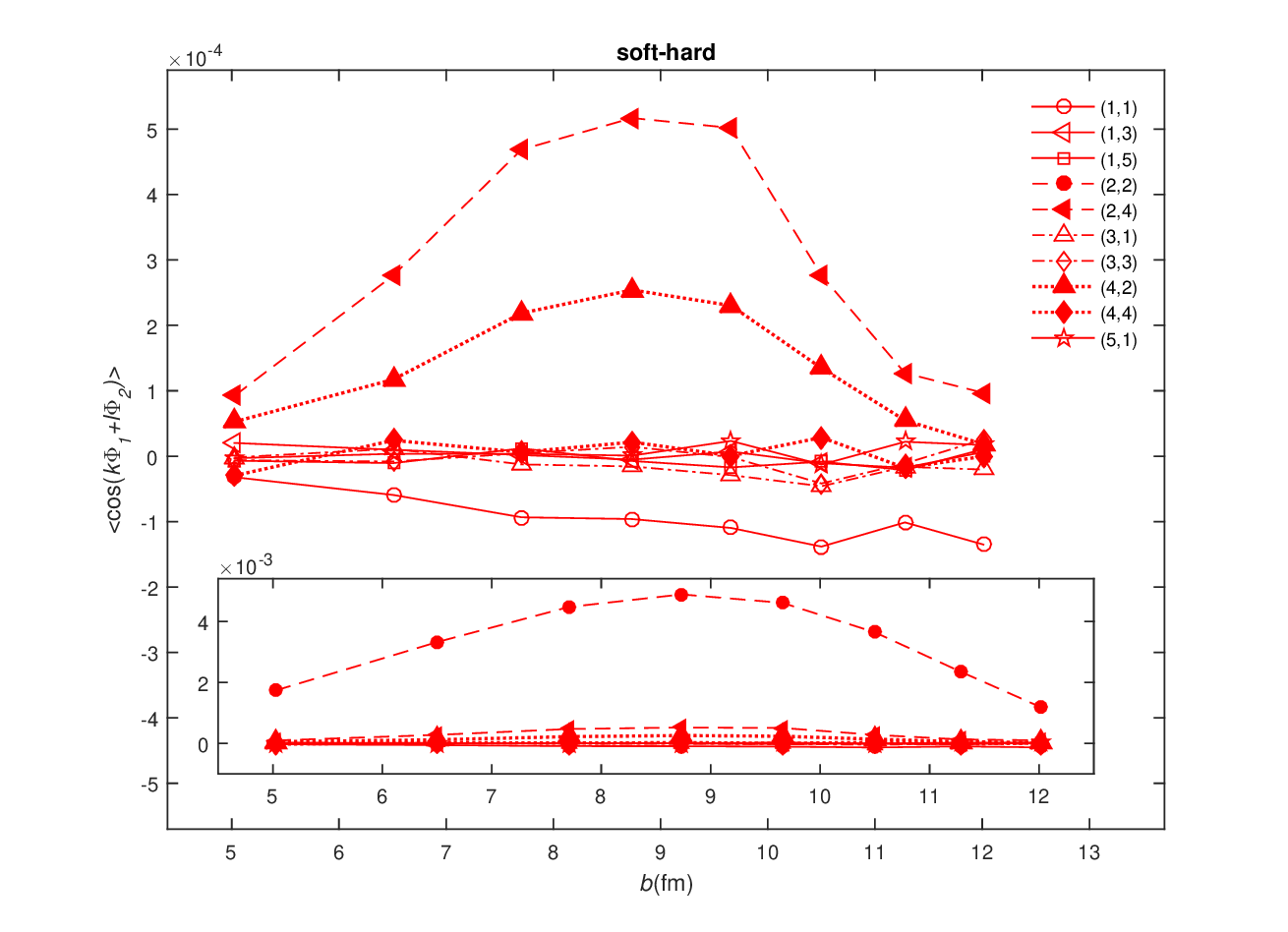}}
\hspace{1in}
\subfigure{ 
\includegraphics[width=0.5\textwidth]{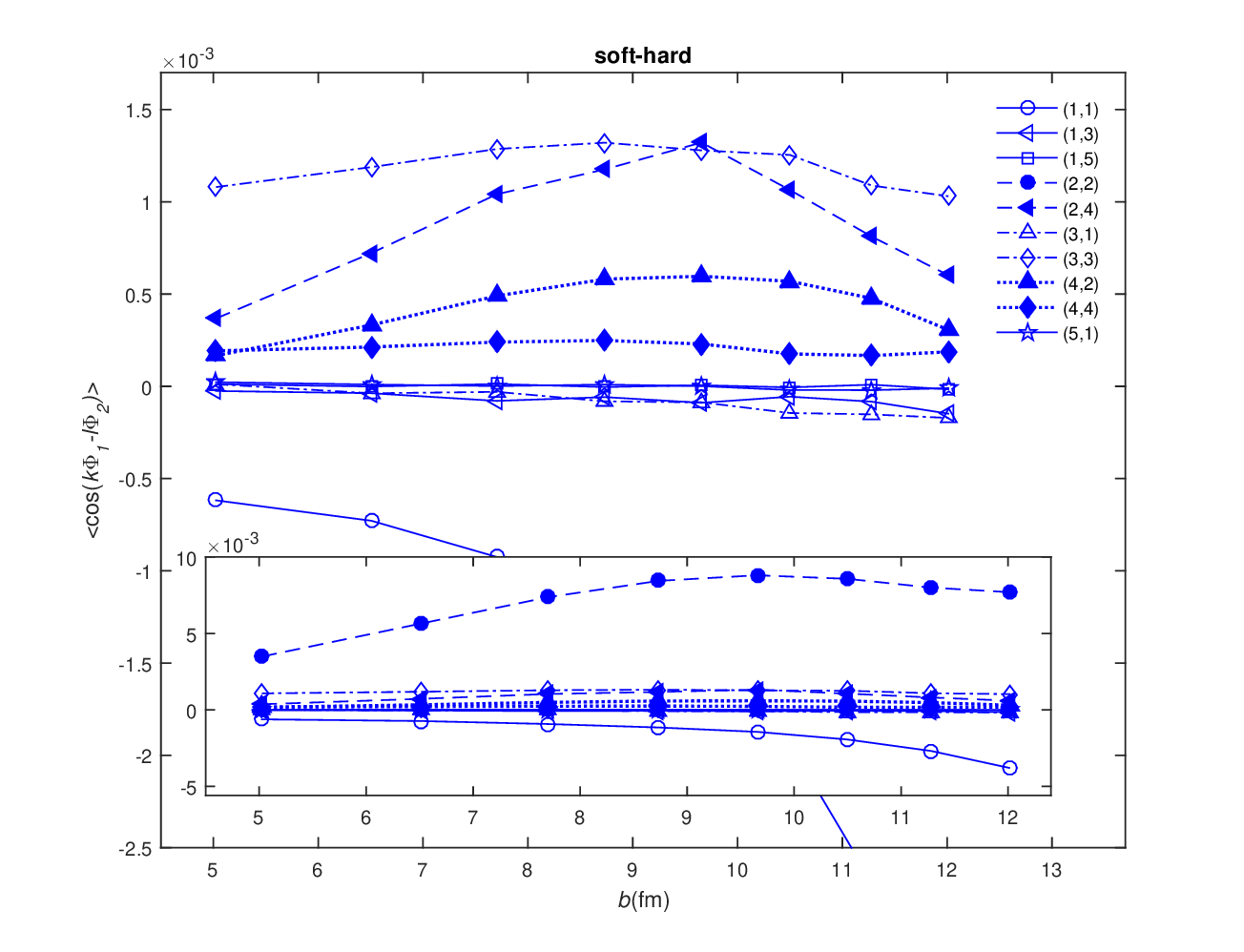}}
\caption{(Color online) Centrality dependence of $\langle\cos(k\Phi_1\pm l\Phi_2)\rangle$ for different combinations of $(k,l)$ in Eq. (\ref{dis2}) for soft-hard particle correlation function  in Eq. (\ref{dis3}).
The soft and hard particles are the same as in Fig.\ref{fig1} and Fig.\ref{fig2}  respectively. The top panel is for  $\langle\cos(k\Phi_1+ l\Phi_2)\rangle$, and the bottom panel is for $\langle\cos(k\Phi_1- l\Phi_2)\rangle$. }
\label{fig4}
\end{figure}

From the generated events, the
flow coefficients $v_n$ for $n=2,4,6$ and the correlation coefficients $\Delta^\pm_{k,l}$
for particle pairs are calculated. For the soft particles, the transverse momentum
averaged flow coefficients are $v_2=0.039, v_4=0.0019, v_6=1.02\times 10^{-4}$
while $v_1\simeq v_3\simeq 0$ for centrality $30-40\% $.
The results for $\langle\cos(k\Phi_1\pm l\Phi_2)\rangle=\Delta^{\pm}_{k,l}/\Delta_{0,0}$ at the same colliding centrality are tabulated in
TABLE I. Comparing $\Delta^\pm_{k,l}/\Delta_{0,0}$ and $v_kv_l$, one can see clearly that the
magnitude of the  later is much smaller
than the former, indicating that the factorization of $\rho_2(\Phi_1, \Phi_2)$ into product of $\rho_1$
is not satisfied. $\langle\cos(k\Phi_1\pm l\Phi_2)\rangle=\Delta^{\pm}_{k,l}/\Delta_{0,0}$ for soft particles at other colliding centralities
are shown in Fig. \ref{fig2} as functions of the impact parameter $b$.

\begin{table}
\center\begin{tabular}{|c|c|c|c|c|}
\hline\hline
$(k, l)$ & $\langle \cos(k\Phi_1-l\Phi_2)\rangle$ & $\langle \cos(k\Phi_1+l\Phi_2)\rangle$\\ \hline
(1, 1) & 8.38$\times 10^{-5}$ & $-6.26\times 10^{-6}$\\ \hline
(1, 3) & $1.49\times 10^{-5}$ & $-3.61\times 10^{-6}$\\ \hline
(2, 2) & $2.36\times 10^{-3}$ & 1.51$\times 10^{-3}$\\  \hline
(1, 5) & $-1.18\times 10^{-6}$ & $-1.32\times 10^{-6}$\\ \hline
(2, 4) & 1.64$\times 10^{-4}$ &$7.61\times 10^{-5}$\\  \hline
(3, 3) & $1.87\times 10^{-4}$ & $-1.20\times 10^{-6}$ \\ \hline\hline
\end{tabular}
\caption{$\langle\cos(k\Phi_1\pm l\Phi_2)\rangle$ for $k+l\leq 6$ in Eq. (\ref{dis2})
for all final state soft hadrons with $|y|<1.0$ and $p_T<1$ GeV/$c$ for Au+Au collisions at $\sqrt{s_{NN}}$
=200 GeV with centrality 30-40\%.  }
\end{table}

 If all the selected particles are produced from jets or within the high $p_T$ region, one
 can obtain expressions and exchange symmetries for the two-particle distributions in the same
 form as for the soft particles, considering the geometrical symmetry of the colliding system.
 For this case, however, two-particle distribution can never be expressed as product of two one-particle   distributions, since jets are produced in pairs almost back to back and particles in one jet are
correlated.  The azimuthal asymmetry for high $p_T$ particles comes from
 the interaction of jets and the produced medium, or in other words by jet quenching \cite{quench}.
 When the number of jets is huge in almost every event,
 the back-to-back correlation  plays an unimportant role, then the factorization
 may be valid approximately.

  With the events generated with AMPT, the flow  coefficients $v_{2n}$ for $n=1$, 2 and 3 and $\Delta^\pm_{k,l}$ in the Fourier expansion of the two-particle  distribution for hard particles can be calculated, as for the soft particles considered in the above. We get  $v_2=0.115$, $v_4=0.0107$,
 and $v_6=1.37\times 10^{-3}$ for hard particles at colliding centrality $30-40\%$. The corresponding values of
$\langle\cos(k\Phi_1\pm l\Phi_2)\rangle=\Delta^{\pm}_{k,l}/\Delta_{0,0}$ at the same centrality
 are tabulated in TABLE II. Values of $\langle\cos(k\Phi_1\pm l\Phi_2)\rangle$ for hard particles are much larger in magnitude than products of $v_n$, as for soft particles. Also
 $\langle\cos(k\Phi_1\pm l\Phi_2)\rangle$  for hard particles are much larger
 than those for the soft particles discussed in the above. $\langle\cos(k\Phi_1\pm l\Phi_2)\rangle=\Delta^{\pm}_{k,l}/\Delta_{0,0}$ for hard particles at other colliding centralities are shown in Fig. \ref{fig3}.

 \begin{table}
\center\begin{tabular}{|c|c|c|c|c|}
\hline\hline
$(k, l)$ & $\langle \cos(k\Phi_1-l\Phi_2)\rangle$ & $\langle \cos(k\Phi_1+l\Phi_2)\rangle$\\ \hline
(1, 1) & $-1.01\times 10^{-3}$ & $-6.43\times 10^{-4}$\\ \hline
(1, 3) & $3.88\times 10^{-6}$ & $-1.66\times 10^{-4}$\\ \hline
(2, 2) & $2.38\times 10^{-2}$ & $1.32\times 10^{-2}$\\  \hline
(1, 5) & $-2.70\times 10^{-4}$ & $-8.75\times 10^{-6}$\\ \hline
(2, 4) & $3.12\times 10^{-3}$ &$1.15\times 10^{-3}$\\  \hline
(3, 3) & $8.53\times 10^{-3}$ & $2.37\times 10^{-4}$\\  \hline\hline
\end{tabular}
\caption{$\langle\cos(k\Phi_1\pm l\Phi_2)\rangle$ for $k+l\leq 6$ in Eq. (\ref{dis2})
for all final state hard hadrons in the $p_T$ larger than 2 GeV/$c$ for Au+Au collisions at $\sqrt{s_{NN}}$
=200 GeV with centrality 30-40\%.}
\end{table}

\begin{figure}[tbhp]  
\includegraphics[width=0.43\textwidth]{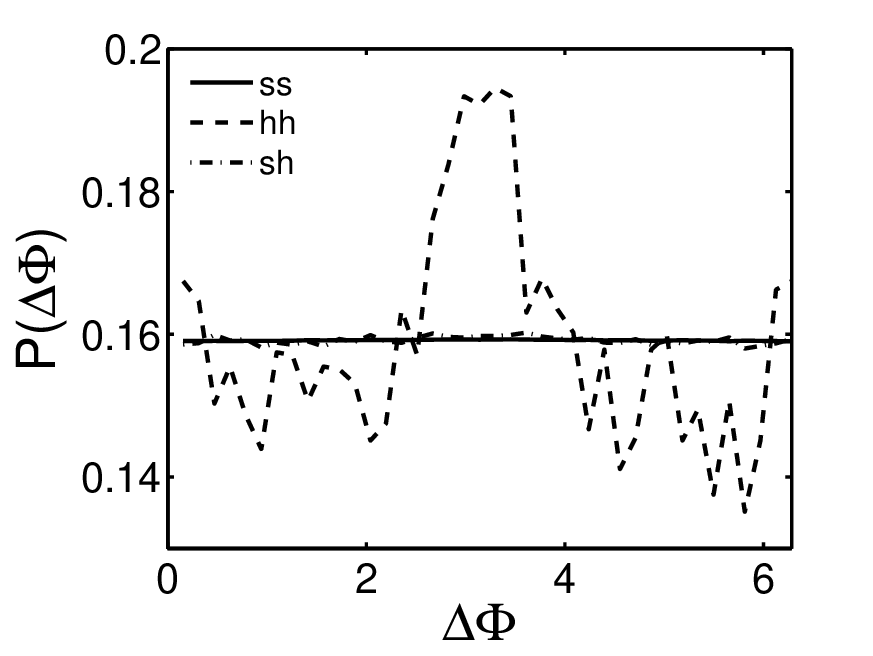}
\caption{$\Delta\Phi$ distributions for soft-soft, hard-hard and soft-hard particle pairs in Au+Au
collisions at $\sqrt{s_{NN}}=200 {\rm GeV}$ at centrality 30-40\%.}
\end{figure}

With the choice of soft and hard particles as in the above,
 the coefficients $\Delta^{\pm}_{k,l}$ for soft-hard correlation in Eq. (\ref{dis3})
 are calculated from the events generated by using AMPT, as used in the above. The results for $\langle\cos(k\Phi_1\pm l\Phi_2)\rangle=\Delta^{\pm}_{k,l}/\Delta_{0,0}$ for centrality 30-40\%
 are tabulated in TABLE \ III, and are shown in Fig. \ref{fig4} for other centralities.
\begin{table}[tbhp]
\center\begin{tabular}{|c|c|c|c|c|}
\hline\hline
$(k, l)$ & $\langle \cos(k\Phi_1-l\Phi_2)\rangle$ & $\langle \cos(k\Phi_1+l\Phi_2)\rangle$\\ \hline
(1, 1) & $-9,25\times 10^{-4}$ & $9.58\times 10^{-5}$\\ \hline
(1, 3) & $-7.58\times 10^{-5}$ & $5.23\times 10^{-6}$\\ \hline
(1, 5) & 5.47$\times 10^{-6}$ & $9.22\times 10^{-7}$\\ \hline
(2, 2) & $7.38\times 10^{-3}$ & $4.47\times 10^{-3}$\\  \hline
(2, 4) & $1.04\times 10^{-3}$ &$4.68\times 10^{-3}$\\  \hline
(3, 1) & $-2.85\times 10^{-5}$ & $-1.38\times 10^{-5}$\\ \hline
(3, 3) & $1.29\times 10^{-3}$ & $9.92\times 10^{-7}$\\  \hline
(4, 2) & $4.89\times 10^{-4}$ & $2.14\times 10^{-4}$\\ \hline
(5, 1) & 7.86$\times 10^{-6}$ & 7.01$\times 10^{-6}$\\ \hline\hline
\end{tabular}
\caption{Averages of $\langle\cos(k\Phi_1\pm l\Phi_2)\rangle$ for $k+l\leq 6$
for soft-hard particle correlation function  in Eq. (\ref{dis3}).
The soft and hard particles are the same as in TABLE I and II, respectively.}
\end{table}

To have a visual comparison among  the two-particle distributions for the soft-soft, hard-hard and
soft-hard particle pairs, one can plot the $\Delta\Phi$ distributions
$P(\Delta\Phi)$ in the same figure for the three sets of
particle pairs. The results are shown in Fig. 5. Because of the fact that $\Delta^-_{k,k}$ for
soft-soft and soft-hard pairs are very small compared to $\Delta^-_{0,0}$, the $\Delta\Phi$ distributions
for those pairs are quite flat. For hard-hard particle pairs, it is a quite different case. A peak can
be seen at $\Delta\Phi=\pi$ in the distribution. This is not surprising, because when a trigger particle
is found with a high $p_T$, it is much more possible that hard particles appear in the opposite
direction in $\phi$, because jets are produced almost back-to-back in heavy ion collisions. Such a
correlation can survive through the averaging process over many collision events.
The asymmetry of $P(\Delta\Phi)$ is not as obvious as observed in \cite{rhic1,rhic2},
because  in this paper the transverse momenta for the triggers and the associated
 particles are much smaller and there are flat contributions from soft particles.

\section{Summary}
In this paper, the general form of two-particle azimuthal distribution is studied
for high energy heavy ion collisions. New variables are suggested for describing the collective behavior
in the final state of the collisions, and the azimuthal correlation function is re-expressed
in terms of those collective variables. Connection between those variables and the well-studied
flow parameters is discussed. Some numerical results on the variables are
presented for Au+Au collisions at $\sqrt{s_{NN}}=200{\rm GeV}/c$ at different centralities from an event generator AMPT.

\begin{acknowledgments}
This work was supported in part by the Ministry of Science and Technology
of China under 973 Grant 2015CB56901, by National Natural Science Foundation of China under Grant Nos. 11435004 and 11375069, and by the Programme of Introducing Talents of Discipline to Universities (B08033).
\end{acknowledgments}

\end{document}